\documentclass[amsmath,amssymb,preprint]{revtex4}
\usepackage{graphicx}
\usepackage{dcolumn}
\usepackage{bm}

\begin{document}

\title{Tunable Rashba spin-orbit interaction at oxide interfaces}
\author{A.D. Caviglia$^{1}$, M. Gabay$^{2}$, S. Gariglio$^{1}$, N. Reyren$^{1}$, C. Cancellieri$^{1}$, J.-M. Triscone$^{1}$}
\affiliation{ $^{1}$D\'epartement de Physique de la Mati\`ere Condens\'ee, University of Geneva, 24 Quai Ernest-Ansermet, 1211 Gen\`eve 4, Switzerland}
\affiliation{ $^{2}$Laboratoire de Physique des Solides, Bat 510, Universit\'e Paris-Sud 11, Centre d'Orsay, 91405 Orsay Cedex, France}
\date{\today}

\begin{abstract}
The quasi-two-dimensional electron gas found at the LaAlO$_{3}$/SrTiO$_{3}$ interface offers exciting new functionalities, such as tunable superconductivity, and has been proposed as a new nanoelectronics fabrication platform. Here we lay out a new example of an electronic property arising from the interfacial breaking of inversion symmetry, namely a large Rashba spin-orbit interaction, whose magnitude can be modulated by the application of an external electric field. By means of magnetotransport experiments we explore the evolution of the spin-orbit coupling across the phase diagram of the system. We uncover a steep rise in Rashba interaction occurring around the doping level where a quantum critical point separates the insulating and superconducting ground states of the system.
\end{abstract}

\maketitle

One of the major quests of modern electronics is the search for new functionalities in solid state nanoscale devices. In that respect, high hopes have been placed on spintronics, where information is processed by manipulating the electrons spin in addition to their charge \cite{zutic:323}. The Spin Field Effect Transistor is certainly the paradigm of this new approach \cite{datta:665}. In this device the Rashba spin-orbit coupling \cite{0022-3719-17-33-015} is used to control the spin precession in a two dimensional electron gas confined in conventional semiconductors heterostructures \cite{PhysRevLett.68.106, PhysRevLett.78.1335, PhysRevLett.90.076807}. Another interesting and potentially rewarding strategy is to develop spintronic devices based on interfaces between complex oxides, where new and unusual electronic phases are promoted \cite{Ohtomo:2004yq, S.Thiel09292006, Brinkman:2007zr, N.Reyren08312007}. In these systems, the Rashba interaction, arising from the breaking of structural inversion symmetry, can be substantial and play a critical role in controlling interfacial electronic states absent in the constituent materials. It has recently been shown that the ground state of the metallic interface between the band insulators LaAlO$_{3}$ and SrTiO$_{3}$ can be driven through a quantum phase transition from an insulating to a superconducting state \cite{Caviglia:2008dq}. Here we show that a strong Rashba spin-orbit interaction is present in this system and that its magnitude can be tuned with an external electric field. Remarkably, a steep rise of the Rashba coupling occurs across the quantum critical point separating the insulating and superconducting ground states of the system \cite{Caviglia:2008dq}. Furthermore, starting from small values at low carrier density, the electron g-factor undergoes a sizable increase, underscoring the concomitant evolution of spin dynamics with doping. The correlation between the critical temperature and the spin-orbit coupling is suggestive of an unconventional superconducting order parameter at the LaAlO$_{3}$/SrTiO$_{3}$ interface.

It has been recently demonstrated by transport experiments \cite{reyren:112506} and conductive atomic force microscopy \cite{Basletic:2008rw, copie:216804} that the electron gas present in LaAlO$_{3}$/SrTiO$_{3}$ heterostructures grown using appropriate conditions (see Supplementary Information) is confined within a few nanometers from the interface. This structural configuration breaks inversion symmetry and, as a result, the electron gas confined in the vicinity of a polar interface \cite{Ohtomo:2004yq} will experience a strong electric field directed perpendicular to the conduction plane. To provide an accurate representation of this internal electric field, the large local polarization of SrTiO$_{3}$ caused by its massive, electric field dependent, low temperature permittivity ($\epsilon_{r}> 10^{4}$) has to be considered \cite{copie:216804}.

A new class of physical phenomena occurring because of the presence of this effective electric field are captured by the Rashba hamiltonian\cite{0022-3719-17-33-015}:
\begin{equation}
H_{\text{R}}=\alpha(\hat{n}\times\vec{k})\cdot\vec{S}
\end{equation}
where $\vec{S}$ are the Pauli matrices, $\vec{k}$ is the electron wave-vector and $\hat{n}$ is a unit vector perpendicular to the interface. This hamiltonian describes the coupling of the electrons spin to an internal magnetic field $\propto\hat{n}\times\vec{k}$, experienced in their rest frame, which is perpendicular to their wave-vector and lies in the plane of the interface. One important consequence of this interaction is that the dispersion relation of the electrons divides into two branches separated at the Fermi surface by a spin splitting $\Delta=2\alpha k_{\text{F}}$, $k_{\text{F}}$ being the Fermi wave-vector and $\alpha$ the strength of the spin-orbit coupling. Perhaps the most appealing feature of this interaction is that its coupling constant is related to the electric field experienced by the electrons and can be therefore tuned by applying an external gate voltage \cite{PhysRevLett.78.1335, PhysRevLett.90.076807}. Aiming to explore this phenomenon in LaAlO$_{3}$/SrTiO$_{3}$ interfaces, we fabricated field effect devices as discussed in the Supplementary Information. In our field effect experiments the modulation of the total electric field experienced by the electron gas is particularly effective thanks to the special dielectric properties of SrTiO$_{3}$.

The magnetic field dependence of the conductance underscores the intriguing coupling between spin dynamics and transport.
\begin{figure}
\includegraphics[scale=0.3]{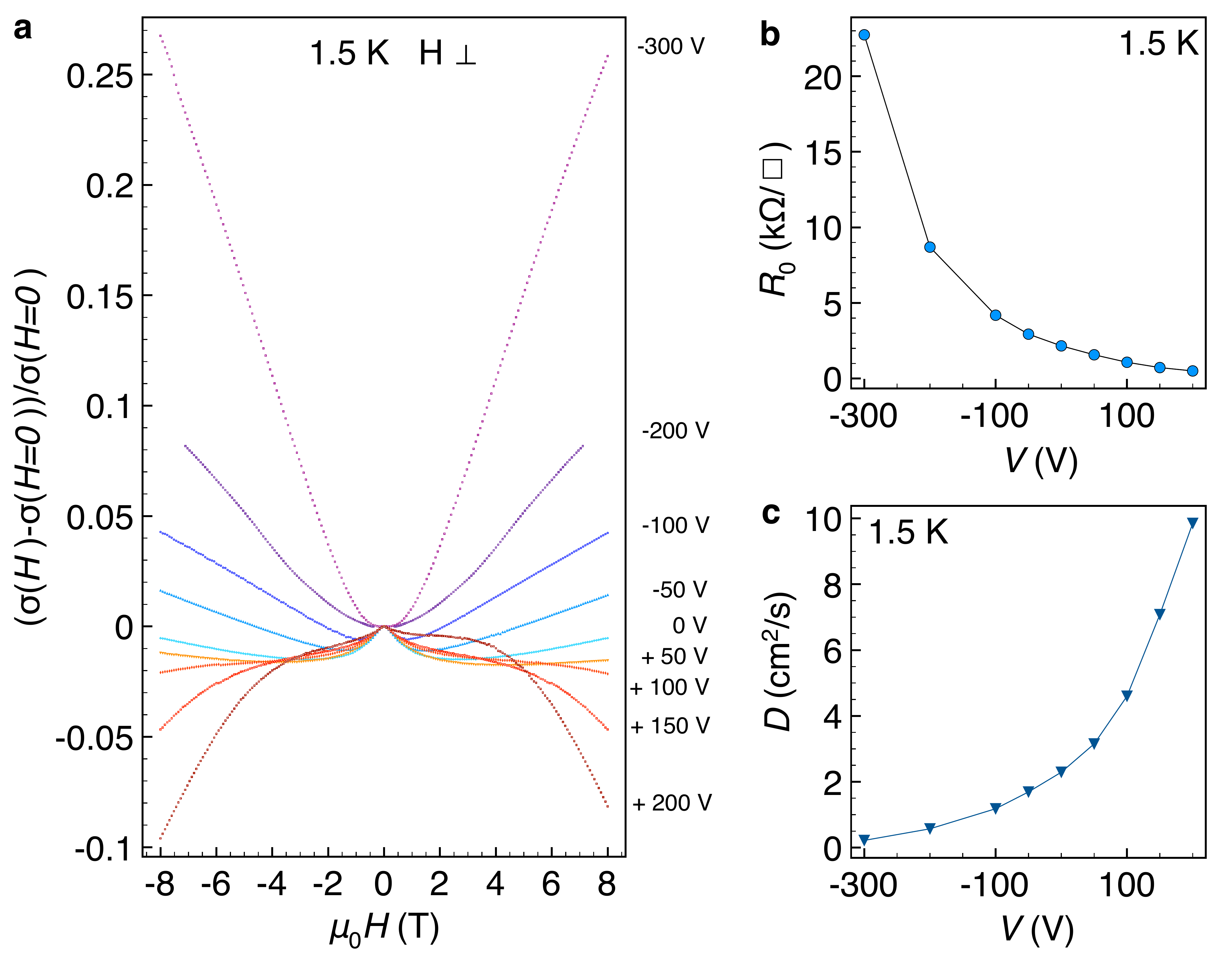}
\caption{\label{fig:mr} Modulation of the transport properties of the LaAlO$_{3}$/SrTiO$_{3}$ interface under electric and magnetic fields. (a) Magnetoconductance $[\sigma(H)-\sigma(H=0)]/\sigma(H=0)$ ($\sigma$ being the sheet conductance, and $H$ the applied magnetic field) measured at 1.5\,K in perpendicular magnetic field for different applied gate voltages. (b) Sheet resistance ($R_{0}$) modulation resulting from the field effect measured at 1.5\,K. (c) Field effect modulation of the diffusion coefficient $D$ estimated at 1.5\,K.}
\end{figure}
Fig. \ref{fig:mr}a shows the magnetoconductance $[\sigma(H)-\sigma(H=0)]/\sigma(H=0)$ ($\sigma$ being the sheet conductance and $H$ the applied magnetic field), measured in a magnetic field applied perpendicular to the LaAlO$_{3}$/SrTiO$_{3}$ interface at a temperature $T= 1.5$\,K for gate voltages $V$ between -300\,V and +200\,V. Measurements performed in a parallel field configuration are presented and discussed in the Supplementary Information. The magnetoconductance measurements are carried out using a standard four points DC technique. As shown in Fig. \ref{fig:mr}a, for large negative gate voltages we observe a large positive magnetoconductance that exceeds +25\% at 8\,T and -300\,V. As we increase the voltage ($V>-200$\,V), a low field regime characterized by a negative magnetoconductance appears. Increasing the gate voltage further, we observe that the negative magnetoconductance regime widens out. For the largest applied electric field, we observe that the magnetoconductance remains negative up to the largest accessible magnetic field (8\,T). This behaviour has been observed in several samples. Similar modulations of magnetoconductance have already been observed in metallic thin films \cite{Bergmann:1984db} and semiconductor heterostructures \cite{PhysRevLett.78.1335, PhysRevLett.90.076807}. Here, the large electrostatic tunability of the magnetoconductance observed in LaAlO$_{3}$/SrTiO$_{3}$ heterostructures is an explicit example of a transport phenomenon occurring at an oxide interface, never observed in its constituent materials. A possible interpretation of this phenomenon is based on the presence of a strong spin-orbit interaction which counteracts weak localization.

Weak localization is a quantum correction to the conductance observed at low temperatures related to the interference of electron waves diffusing around impurities \cite{Bergmann:1984db}. Neglecting spin effects one finds that this contribution to the conductance is always \textit{negative} and that a perturbation which breaks time reversal invariance such as an external magnetic field increases the conductance. As shown in Fig. \ref{fig:mr}a, this is what is observed in LaAlO$_{3}$/SrTiO$_{3}$ interfaces for large negative bias. In addition, in the presence of spin-orbit interaction, it has been shown both theoretically \cite{PTP.63.707} and experimentally \cite{Bergmann:1984db}, that electron interference will bring about a \textit{positive} contribution to the conductance. The \textit{positive} contribution is the first to be suppressed by an external magnetic field causing an initial decrease in conductance as a function of magnetic field \cite{Bergmann:1984db}. This unexpected behaviour, named weak antilocalization, is observed in LaAlO$_{3}$/SrTiO$_{3}$ heterostructures polarized above a particular gate field.
The spin-orbit relaxation time is therefore an essential ingredient to describe transport in a two-dimensional electron gas in the presence of a strong homogeneous electric field. As previously discussed, the conduction electrons will experience an internal magnetic field which is always perpendicular to their wave vector $\vec{k}$. In a diffusive system this vector will rotate at every scattering event causing rapid fluctuations of the internal magnetic field. These fluctuations will affect the evolution of the spin phase and will define a spin relaxation time $\tau_{\text{so}}$. This is known as the D'yakonov-Perel' (DP) mechanism of spin relaxation \cite{DP}. In this scenario, the Rashba coupling constant $\alpha$ and the spin relaxation time $\tau_{\text{so}}$ are related through
\begin{equation}\label{eq:alpha}
\tau_{\text{so}}=\frac{\hbar^{4}}{4\alpha^{2}m^{2}2D}
\end{equation}
where $m$ is the carrier mass and $D$ the diffusion constant. An additional spin-orbit term pertaining to the DP process is discussed in the Supplementary Information.
A second class of spin relaxation processes, known as the Elliott-Yafet (EY) mechanism, originates from the spin-orbit interaction of the lattice ions with the conduction electrons \cite{PhysRev.96.266, yafet}. This spin relaxation mechanism can become relevant in the presence of strong spin-orbit scattering impurities or whenever the ionic spin-orbit coupling produces a significant correction to the band structure of the material. In the case of SrTiO$_{3}$, band structure calculations show that the Ti 3d conduction bands are notably altered by this correction \cite{PhysRevB.6.4740}. Therefore, in principle both mechanisms can be at play at the LaAlO$_{3}$/SrTiO$_{3}$ interfaces. Nevertheless one can identify the dominant mechanism by studying the dependence of the spin relaxation time on the elastic scattering time $\tau$ \cite{zutic:323}. In the case of the EY mechanism (ionic spin-orbit interaction), the Elliott relation $\tau_{so}\sim \tau/(\Delta g)^{2}$ ($\Delta g$ is the difference between the electrons g-factor in the solid and the one of free electrons) predicts a direct proportionality between the spin relaxation time and the elastic scattering time. In a DP scenario (Rashba spin-orbit interaction) the spin relaxation time should be inversely proportional to the elastic scattering time $\tau_{so}\sim 1/\tau$.

The influence of the Rashba term, Eq. (2), is assessed by measuring the magnetoconductivity in the diffusive regime. In a two-dimensional layer with in-plane spin-orbit relaxation time, immersed in a perpendicular magnetic field $H$, and in the limit $H<H_{\text{so}}=\hbar/4eD\tau_{\text{so}}$, the first order correction to the conductance, $\Delta\sigma$, takes the Maekawa-Fukuyama (MF) form \cite{JPSJ.50.2516} (see Supplementary Information)
\begin{equation*}
\frac{\Delta\sigma(H)}{\sigma_{0}}=\Psi\left(\frac{H}{H_{\text{i}}+H_{\text{so}}}\right)+\frac{1}{2\sqrt{1-\gamma^{2}}}\Psi\left(\frac{H}{H_{\text{i}}+H_{\text{so}}\left(1+\sqrt{1-\gamma^{2}}\right)}\right)
\end{equation*}
\begin{equation}\label{eq:MF}
-\frac{1}{2\sqrt{1-\gamma^{2}}}\Psi\left(\frac{H}{H_{\text{i}}+H_{\text{so}}\left(1-\sqrt{1-\gamma^{2}}\right)}\right)
\end{equation}
The function $\Psi$ is defined as
\begin{equation}
\Psi(x)=\ln(x)+\psi\left(\frac{1}{2}+\frac{1}{x}\right)
\end{equation}
where $\psi(x)$ is the digamma function and $\sigma_{0}=e^{2}/\pi h$ is a universal value of conductance. The parameters of the theory are the inelastic field $H_{\text{i}}=\hbar/4eD\tau_{\text{i}}$, $H_{\text{so}}$ and the electrons g-factor $g$ which enters into the Zeeman correction $\gamma=g\mu_{B}H/4eDH_{\text{so}}$. $\mu_{B}$ is the Bohr magneton and $\tau_{\text{i}}$ is the inelastic scattering time. Since we perform our experiments at 1.5\,K, which is at least 5 times the maximum superconducting critical temperature, we can neglect superconducting fluctuations. In a magnetotransport experiment we can then quantify the two relevant time scales of the problem, namely $\tau_{\text{so}}$ and $\tau_{\text{i}}$.

\begin{figure}
\includegraphics[scale=0.3]{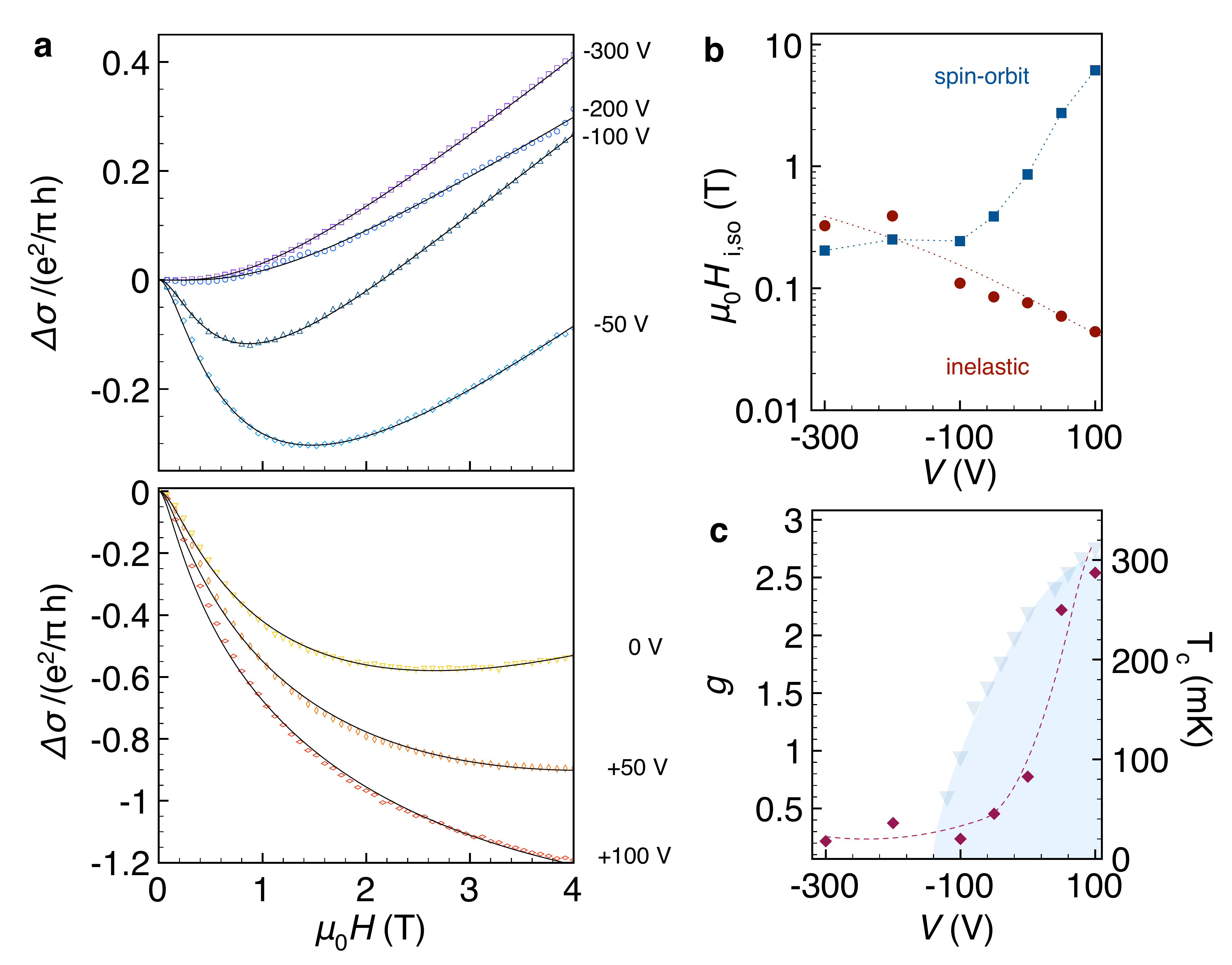}
\caption{\label{fig:fit} Analysis of the magnetoconductivity of the LaAlO$_{3}$/SrTiO$_{3}$ interface. (a) Best fits according to the Maekawa-Fukuyama theory of the variation of conductance $\Delta\sigma$, normalized with respect to $e^{2}/\pi h$, for different gate voltages. (b) Gate voltage dependence of the fitting parameters $H_{\text{i}}$ (red dots) and $H_{\text{so}}$ (blue squares). The lines are a guide to the eye. (c) Left axis, purple diamonds: gate voltage dependence of the electrons g-factor $g$. The line is a guide to the eye. Right axis, blue triangles: superconducting critical temperature $T_{c}$ as a function of gate voltage for the same sample.}
\end{figure}

The MF theory has been used to fit the experimental data of Fig. \ref{fig:mr}a in terms of variation of conductance with respect to $e^{2}/\pi h\simeq1.2 \cdot 10^{-5}$\,S. Since the effective mass (the elastic scattering time) is one to two orders of magnitude larger (smaller) than the corresponding quantities for typical semiconductors, the diffusive regime holds for fields up to 4\,T. As the MF theory is based on a perturbative expansion, we have also checked that the magnetoresistance and magnetoconductance are still equal in absolute value up to 4\,T. For the range of fields and gate voltages (up to 100 V \footnote{As discussed in Supplementary Information, for $V>100$\,V the MF theory alone is not adequate to fit the experimental data, indicating that additional contributions to the magnetoconductance need to be taken into account.}) that we analyzed , weak localization corrections dominate Coulomb interaction contributions. We also explicitly verified that superconducting fluctuations did not contribute significantly to the magnetoconductance. The best fits are presented in Fig. \ref{fig:fit}a, where we observe a remarkable agreement between theory and experiments. This analysis allows us to trace the electric field dependence of the parameters $H_{\text{i,so}}$, presented in Fig. \ref{fig:fit}b, and $\gamma$. Analyses of the magnetoconductance performed using expressions derived by A. Punnoose \cite{punnoose:252113} provide the same evolution of the characteristic fields.
To extract from these parameters the relaxation times $\tau_{\text{i,so}}$ and the electrons g-factor, we need to determine the electric field dependence of the diffusion coefficient. For this purpose we measured the electric field modulation of the sheet carrier concentration $n_{\text{2D}}$ by means of Hall effect and by capacitance measurements \cite{Caviglia:2008dq} (see Supplementary Information). An estimate of the Fermi velocity $v_{\text{F}}$ and of the elastic scattering time using a parabolic dispersion relation with an effective mass $m^{*}=3m_{e}$ \cite{PhysRevB.6.4740}, ($m_{e}$ is the bare electron mass) and data collected at the temperature $T=1.5$\,K, allows the diffusion coefficient to be derived as $D=v_{\text{F}}^{2}\tau/2$. $D$ as a function of $V$ is plotted in Fig. \ref{fig:mr}c.

The gate voltage dependence of the g-factor is presented in Fig. \ref{fig:fit}c. One observes a large increase from a small value, around 0.5 for negative voltages, towards the typical value of 2 for bare electrons at positive voltages. An electric field control of the g-factor has been previously predicted \cite{Ivchenko1997375} and experimentally demonstrated \cite{Salis:2001ye} in semiconductor heterostructures. These studies have correlated the modulation of the g-factor to a modification of the spatial distribution of the electronic wave functions driven by an external electric field. Alternatively, this variation could reveal correlation effects, either the formation of incoherent Cooper pairs or the presence of a pseudogap, as is seen in high-$T_{c}$ superconductors \cite{lee:17}. In LaAlO$_{3}$/SrTiO$_{3}$ interfaces we notice that the sharp increase of the g-factor appears in the vicinity of the quantum critical point as we approach the superconducting region of the phase diagram. Additional investigations will be necessary to further our understanding.
We now turn to the issue of the gate voltage dependence of the parameters $H_{\text{i,so}}$ that will allow us to discern the modulation of spin-orbit coupling brought about by the electric field.
\begin{figure}
\includegraphics[scale=0.3]{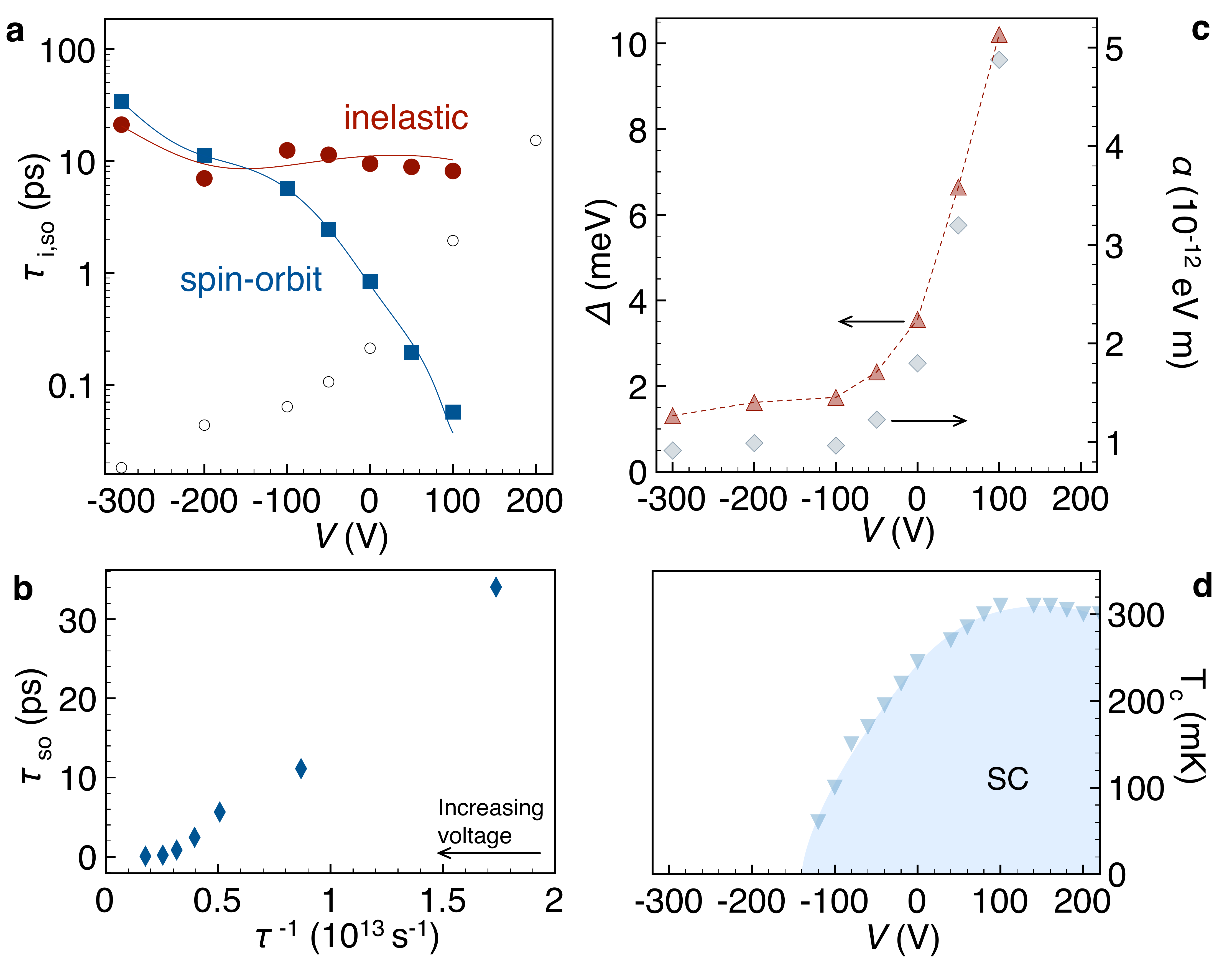}
\caption{\label{fig:length}Rashba control of the LaAlO$_{3}$/SrTiO$_{3}$ interface electronic phase diagram. (a) Inelastic relaxation time $\tau_{\text{i}}$ (red circles) and spin relaxation time $\tau_{\text{so}}$ (blue squares) as a function of gate voltage plotted on a logarithmic time scale. The lines are a guide to the eye. Prediction of the spin relaxation time as a function of gate voltage based on the Elliott relation (open circles). (b) Spin relaxation time vs elastic scattering rate showing consistency with the D'yakonov-Perel' mechanism. (c) Left axis, red triangles: field effect modulation of the Rashba spin splitting $\Delta$. Right axis, grey diamonds: field effect modulation of the Rashba coupling constant $\alpha$. (d) Superconducting critical temperature $T_{c}$ as a function of gate voltage for the same sample. Note that the crossing of the inelastic and spin relaxation times occurs at the quantum critical point.}
\end{figure}

The relaxation times $\tau_{\text{i,so}}$ are plotted against gate voltage in Fig. \ref{fig:length}a. For large negative gate voltages we observe that the inelastic scattering time is shorter than the spin relaxation time, indicating that the effect of the spin-orbit interaction is weak compared with the orbital effect of the magnetic field. In this regime, the quantum correction to the conductivity can be ascribed to weak localization, in agreement with the the observed temperature evolution of the conductivity \cite{Caviglia:2008dq}. Above a critical voltage the spin relaxation time becomes shorter than the inelastic scattering time and decreases sharply, by three orders of magnitude, as the voltage is increased. By contrast, the inelastic scattering time remains fairly constant as we increase the voltage. Here a weak antilocalization regime appears, characterized by a strong spin-orbit interaction. As previously discussed, the nature of the spin-orbit mechanism can be discerned by examining the dependence of the spin relaxation time on the elastic scattering time. In Fig. \ref{fig:length}a we show the gate voltage dependence of the spin relaxation time predicted by the Elliott relation, calculated using the electrons g-factor presented in Fig.  \ref{fig:fit}c. Clearly, the EY mechanism fails to estimate the spin relaxation time by 3 orders of magnitude at -300\,V and its predicted variation with $V$ is opposite to that observed. In fact, as can be seen in Fig. \ref{fig:length}b, the spin relaxation time is proportional to the inverse of the elastic scattering time over a wide voltage range, a clear signature of the DP mechanism characteristic of the Rashba spin-orbit interaction. For $V>$ 50\,V, a deviation from the DP relation is experimentally observed. This deviation coincides with the departure from the collision-dominated regime which occurs as $\tau_{\text{so}}$ becomes of order $\tau$. This evolution points towards a strong spin-orbit coupling where an electron spin may precess through
several cycles before scattering. These observations indicate that the unusually strong and tunable spin-orbit interaction found in LaAlO$_{3}$/SrTiO$_{3}$ heterostructures arises from the interfacial breaking of inversion symmetry. Calculations by Copie et al.\cite{copie:216804} show that in this system a variation of the charge density is accompanied by a large modulation of the interfacial electric field. Moreover, the charge profile of the electron gas in SrTiO$_{3}$ is strongly modified as the carrier density is varied. Ultimately, it is this modification of the asymmetry of the charge profile that strongly affects the evolution of the Rashba coupling.

A remarkable correlation between the onset of strong spin fluctuations and the emergence of superconductivity is evident by comparing Fig. \ref{fig:length}a and d, where we notice that the superconducting dome, measured on the same sample, develops as the spin relaxation time becomes significantly smaller than the inelastic scattering time. This finding suggests that the spin-orbit interaction plays an important role in stabilizing a delocalized phase in two dimensions \cite{Papadakis:1999kb} which condenses into a superconducting state. The gate voltage dependence of the diffusion coefficient previously presented corroborates this interpretation. In Fig. \ref{fig:length}c we can appreciate the sharp increase of the spin-orbit coupling constant $\alpha$, calculated using equation \ref{eq:alpha}, as we move across the quantum critical point and the corresponding rise of the spin splitting $\Delta$. This remarkable correlation between the critical temperature and the intensity of spin fluctuations is suggestive of an unconventional superconducting order parameter at the LaAlO$_{3}$/SrTiO$_{3}$ interface \cite{PhysRevLett.87.037004, PhysRevLett.75.2004}. The large change in spin relaxation across the phase diagram, which is not yet fully understood, may result from a complex dependence of the spin-orbit coupling on band structure properties and charge profile asymmetry and should stimulate further theoretical and experimental investigations (see Supplementary Information). We note that the spin splitting values can be much higher than the superconducting gap (which is of the order of 40\,$\mu$eV at optimal doping) and comparable to the Fermi energy (which is of the order of 20\,meV). Hence the spin-orbit coupling turns out to be an essential ingredient to describe the electronic properties of the LaAlO$_{3}$/SrTiO$_{3}$ interface, both in the normal and superconducting state.

In conclusion, the electric field control of spin coherence at the interface between two complex oxides, LaAlO$_{3}$ and SrTiO$_{3}$, has been demonstrated. Quantum interference effects have been used as a probe of spin dependent transport, uncovering a remarkable correlation between the Rashba spin-orbit coupling and the electronic phase diagram. These observations suggest that a manipulation of the spin-orbit coupling in complex oxides interfaces can potentially drive fundamental changes in their ground states, opening a new pathway toward spintronics. The ability to control the Zeeman and Rashba spin splitting in an oxide system is particularly promising, as fully spin polarized materials can be integrated in functional heterostructures. Moreover this technology can be applied to nanoscale devices, where spin coherence can be manipulated by local electric fields \cite{ChengCen02202009}.

\subsection{Acknowledgements}
We thank D. Jaccard, A.F. Morpurgo and M. Bibes for useful discussions and Marco Lopes for his technical assistance. We acknowledge financial support by the Swiss National Science Foundation through the National Centre of Competence in Research ``Materials with Novel Electronic Properties" MaNEP and Division II, by the European Union through the projects ``Nanoxide" and ``OxIDes", and by the European Science Foundation through the program ``Thin Films for Novel Oxide Devices".
\newpage
\section{Supplementary Information - Field effect devices fabrication and characterisation}

Field effect devices based on LaAlO$_{3}$/SrTiO$_{3}$ conducting interfaces were fabricated by depositing LaAlO$_{3}$ epitaxial films at least 4 unit cells (u.c.) thick \cite{S.Thiel09292006} onto TiO$_{2}$ terminated (001) SrTiO$_{3}$ single crystals. The films were grown by pulsed laser deposition at $\sim 800$\,$^{\circ}$C in $\sim 1\times 10^{-4}$ mbar of O$_{2}$ with a repetition rate of 1 Hz. The fluence of the laser pulses was 0.6 J/cm$^{2}$. The film's growth was monitored \textit{in
situ} using reflection high energy electron diffraction which allowed the thickness to be controlled with sub-unit-cell precision. After growth, each sample was annealed in 200 mbar of  O$_{2}$ at about 600\,$^{\circ}$C for one hour and cooled to room temperature in the same oxygen pressure \cite{0953-8984-21-16-164213}. 500\,$\mu$m wide transport channels for four points measurements were defined by optical lithography \cite{Caviglia:2008dq}. The 0.5\,mm thick SrTiO$_{3}$ substrate was used as the gate dielectric. The metallic gate is a gold film sputtered on the backside of the substrate opposite to the channel area.
The $C(V)$ characteristics of the device were measured using an Agilent 4284A LCR meter. The variation of the carrier concentration $\delta n_{2D}$ between gate voltages $V_{1}$ and $V_{2}$ can then be estimated using the relation
\begin{equation}\label{eq:cv}
\delta n_{2D} = \frac{1}{Se}\int_{V_{1}}^{V_{2}}C(V)\text{d}V
\end{equation}
where $S$ is the area of the gate electrode and $e$ is the elementary charge.
Fig. S1 shows the modulation of the carrier concentration estimated at 1.5\,K from the $C(V)$ characteristics.
In order to obtain reproducible $C(V)$ and $R(V)$ characteristics during the experiment the electric field is first ramped to the highest positive voltage.
\renewcommand{\thefigure}{S\arabic{figure}}

\begin{figure} 
\includegraphics[scale=0.3]{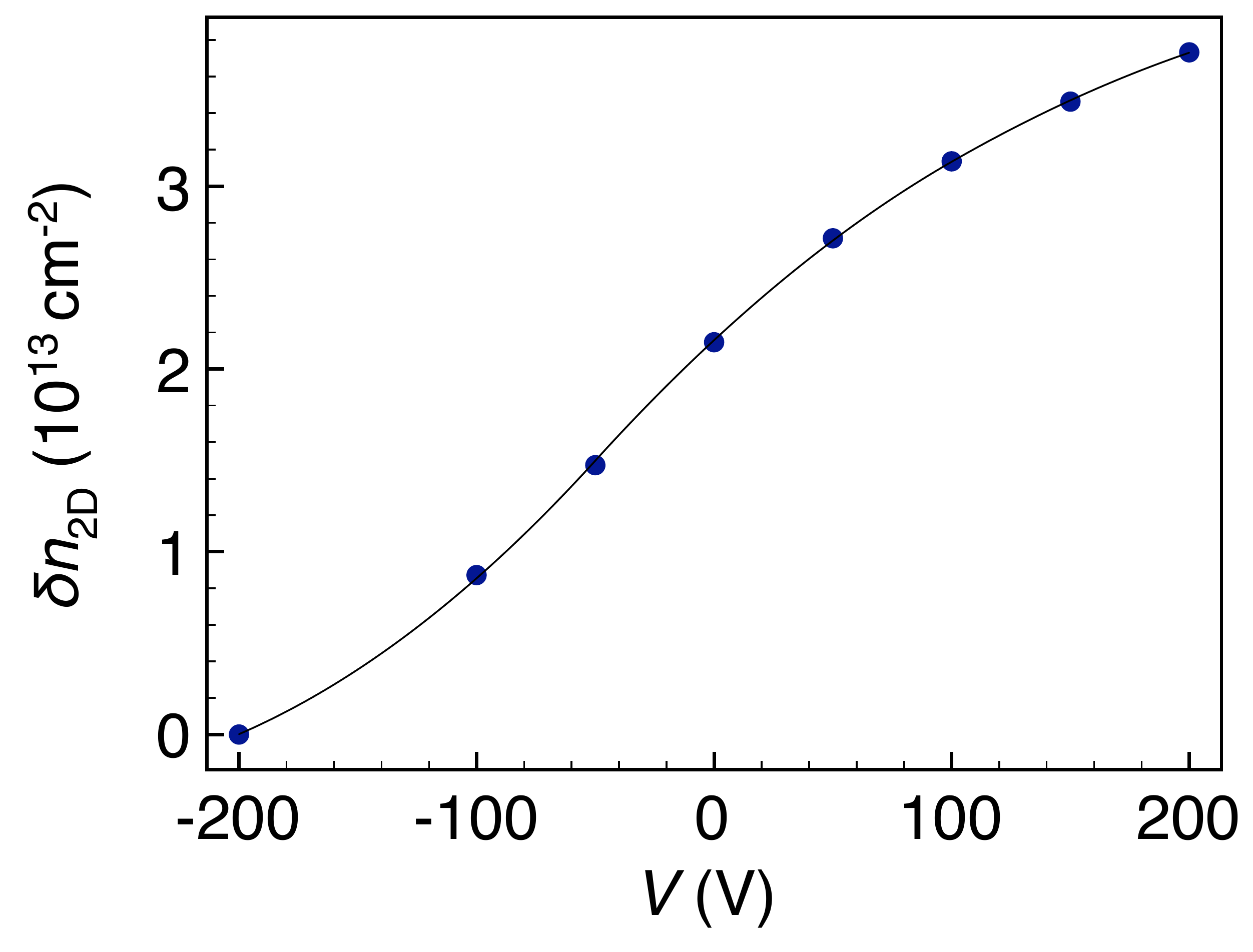}
\caption{\label{fig:mr} Modulation of the carrier concentration with gate voltage for the LaAlO$_{3}$/SrTiO$_{3}$ interface at 1.5\,K estimated using eq. 1 with $V_{1}=-200$\,V and $V_{2}=V$. From Ref. \cite{Caviglia:2008dq}.}
\end{figure}

\section{Transport in perpendicular and parallel magnetic fields}
The magnetoconductance of a disordered electronic system is determined by different contributions that can be classified into two categories: one is interaction (Coulomb) and the other is weak localization (WL). The former class consists of exchange and Hartree terms (showing in Zeeman-type effects) while the latter consists of orbital, Zeeman and spin-orbit terms. Space inversion asymmetry promotes spin-orbit scattering processes of the Dresselhaus \cite{PhysRev.100.580} and Rashba \cite{0022-3719-17-33-015} types. In the diffusive limit, both of these effects contribute to the D'yakonov-Perel' (DP) spin relaxation rate, and hence enter the expression of the WL correction to the magnetoconductivity $\Delta \sigma$ \cite{ilp, PhysRevB.51.16928, PhysRevB.53.3912}.The Dresselhaus term contains two pieces. In a confined geometry, one is proportional to (1- $d^{2}n_{\text{2D}}/2\pi$) \cite{PhysRevB.51.16928, Koralek:2009rc}, where $d$ is the thickness of the electron gas layer. From estimates of $d$ \cite{Basletic:2008rw, reyren:112506} and of $n_{\text{2D}}$, one sees that this part is essentially zero. The other piece enters the expression of the spin-orbit field Hso and adds to the Rashba term. Fitting the data to the analytical expression of $\Delta \sigma$ for fields $H< H_{\text{so}}$ and $H>H_{\text{so}}$ shows that the Rashba contribution dominates. In that case, a closed form of  $\Delta \sigma$ has been derived by Punnoose \cite{punnoose:252113}, which reduces to the Maekawa-Fukuyama expression, in the limit $H< H_{\text{so}}$. Including the Zeeman term allows us to determine $H_{\text{so}}$, the inelastic field $H_{\text{i}}$, and also $g$. We find that the same parameters provide an accurate fit of the data for $H>H_{\text{so}}$. In 2D, in the diffusive regime, both WL and interaction effects yield logarithmic corrections to the magnetoconductance. It is thus difficult, analyzing only the magnetoconductance in perpendicular field, to sort out the different contributions. However,  a magnetic field applied parallel to the conducting 2D layer quenches the orbital motion and, provided the spin-orbit contribution is negligible, leaves only the Coulomb contribution, whose strength depends on the electrons g-factor $g$. In this case, the magnetic field polarizes the spins and leads to a negative contribution to the magnetoconductance. Similarly, if one applies a strong enough perpendicular field, one freezes out the WL correction which reveals the Coulomb contribution. In this field orientation, it is generally accepted that the orbital and spin-orbit corrections dominate for fields that satisfy $g\mu_{\text{B}}H/k_{\text{B}}T<<1$.

Fig. S2 displays magnetoconductance measurements for the parallel and perpendicular field orientations. For gate voltages smaller than 0 V, the Zeeman contribution is small up to 4\,T. One has to recall (see Fig. 2c) that $g$ is small so that, in this concentration range, WL effects dominate for both field orientations. In this regard, we remark that the Maekawa-Fukuyama's theory predicts in the presence of spin-orbit scattering a weak (compared to the perpendicular case) negative parallel magnetoconductance \cite{JPSJ.50.2516}. This  is precisely what we observe for this range of gate voltages as shown in Fig. S2. This observation, together with the absence of hysteresis, suggests that the positive magnetoconductance observed in the perpendicular orientation does not arise from magnetic effects.

Since the parameters that are extracted using Eq. 2 give excellent fits to the experimental data both at low and high perpendicular fields, this indicates that for gate voltages up to 100\,V, Coulomb interactions yield small contributions compared to WL.
We remark (Fig. S2) for high concentrations (150 V,  200 V and 250 V) that in small parallel fields (less then 2 T) a negative magnetoconductance is observed as expected from spin-orbit and Coulomb contributions. The behaviour observed at higher parallel fields (positive magnetoconductance) is unclear at the moment and might be explained by the finite thickness of the electron gas \cite{altsciulerro}. In the high positive voltage range a fit of the experimental data in perpendicular field could not be obtained by using WL contributions only. Since in this doping range the mobility is significantly increased (see Fig. 1c), this could indicate a transition towards a Shubnikov-de Haas regime \cite{sedrakyan:106806}.

\begin{figure}
\includegraphics[scale=0.3]{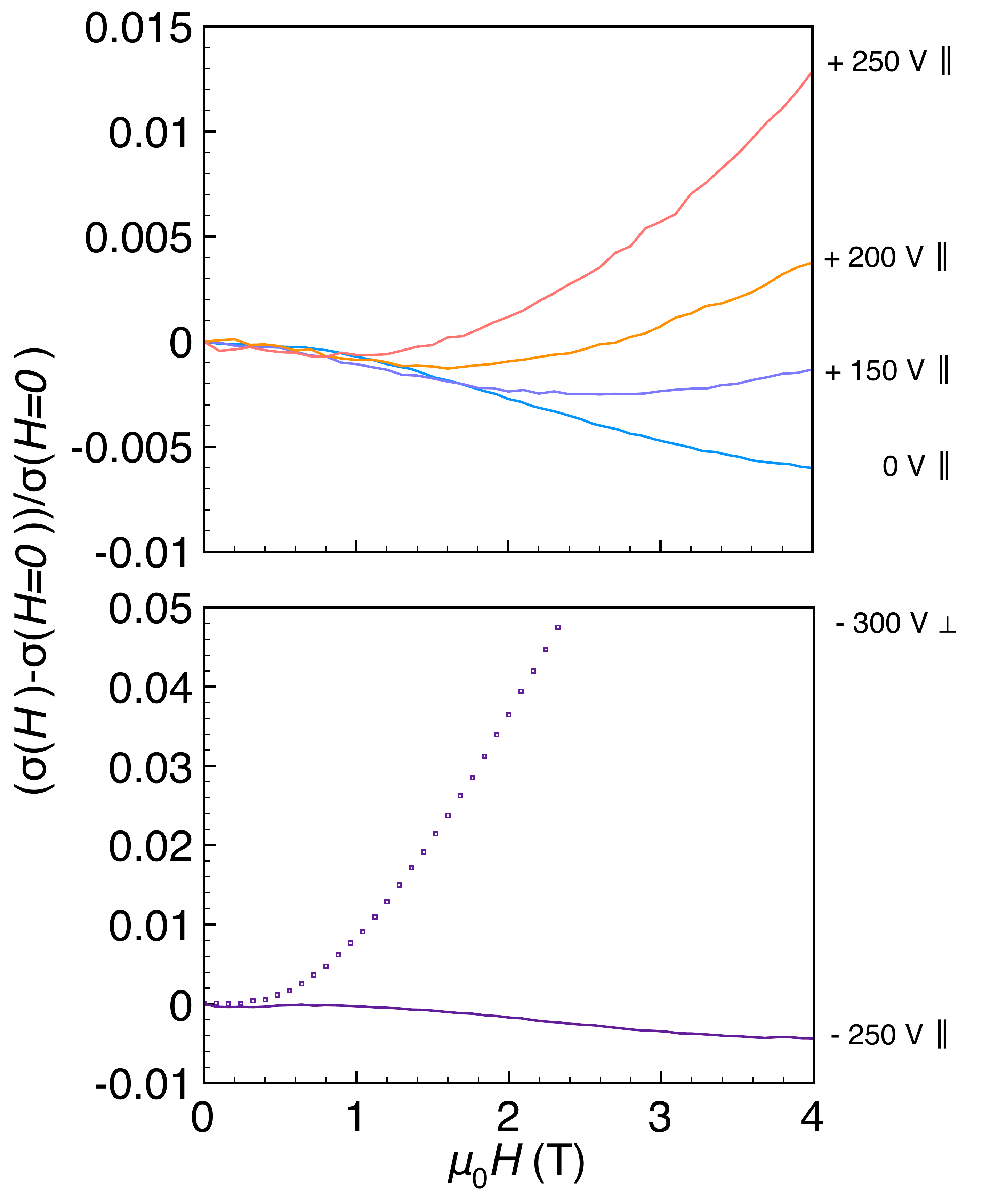}
\caption{\label{fig:mr} Magnetoconductance under electric field at 1.5\,K. with magnetic field applied perpendicular and parallel to the LaAlO$_{3}$/SrTiO$_{3}$ interface.}
\end{figure}

\section{Mechanisms for field effect modulation of spin-orbit coupling in LaAlO$_{3}$/SrTiO$_{3}$ interfaces}
In our experiments on LaAlO$_{3}$/SrTiO$_{3}$ interfaces, we observe a remarkable evolution of the strength of the spin-orbit coupling $\alpha$ in response to the modulation of the gate voltage. In two-dimensional electron gases, $\alpha$ depends on the electric field as well as on the band structure, gap energy and Pauli spin-orbit interaction \cite{PhysRevLett.60.728, winkler, ivchenko}. For SrTiO$_{3}$, according to calculations by Mattheiss \cite{PhysRevB.6.4740}, a Pauli spin-orbit coupling related to Ti ions already exists and produces a significant correction to the conduction band structure. Also, in LaAlO$_{3}$/SrTiO$_{3}$ interfaces, the asymmetry of the charge profile can evolve dramatically with the modulation of the external gate voltage due to the particular dielectric properties of SrTiO$_{3}$ \cite{copie:216804}. Thus, the observed large variation of the spin-orbit coupling could be a result of a complex interplay between Pauli spin-orbit interaction, charge profile asymmetry, and band filling that are modified by the gate voltage.

\end{document}